# Prefrontal cortex functional connectivity based on simultaneous record of electrical and hemodynamic responses associated with mental stress

**FARES AL-SHARGIE,**[1]*

[1]*Department of Electrical Engineering, American University of Sharjah, Sharjah, United Arab Emirates*
[2]*Universiti Teknologi PETRONAS, Centre of Intelligent Signal and Imaging Research, Department of Electrical and Electronic Engineering, 32610 Bandar Seri Iskandar, Perak, Malaysia*
*\*fyahya@aus.edu*

**Abstract:** This paper investigates prefrontal cortex (PFC) functional connectivity based on synchronized electrical and hemodynamic responses associated with mental stress. The electrical response was based on alpha rhythmic of Electroencephalography (EEG) signals and the hemodynamic responses were based on the mean concentrations of oxygenated and deoxygenated hemoglobin measured using functional Near-Infrared Spectroscopy (fNIRS).
The aim is to explore the effects of stress on the inter and intra hemispheric PFC functional connectivity at narrow and wide frequency bands with 8- 13 Hz in EEG and 0.009-0.1Hz in fNIRS signals. The results demonstrated significantly reduce in the functional connectivity on the dorsolateral PFC within the inter and intra hemispheric PFC areas based in EEG alpha rhythmic and fNIRS oxygenated and deoxygenated hemoglobin. The statistical analysis further demonstrated right dorsolateral dominant to mental stress.

## 1. Introduction

Stress is one of the major health problem worldwide. Long term exposure to workload related-stress activities has been associated with adverse mental health problems [1]. Stress contributes to risk factors for neuropsychiatric disorders such as bipolar disorders, schizophrenia, anxiety and depression [2-4]. Stress disrupts creativity, problem solving, decision making, working memory and other prefrontal cortex (PFC)-dependent activities [5-8]. Studies have demonstrated that, exposure to acute or uncontrollable stress increased the catecholamine release in the prefrontal cortex resulted in reducing the neuronal firing and impaired cognitive abilities in both animal and human subjects [9,10]. Chronic stress has also been shown to weakness PFC functional connectivity and reshape its structure [6,11]. PFC topographically organized to the dorsolateral PFC, ventrolateral PFC and the frontopolar area FPA. The dorsolateral PFC (DLPFC) guides thoughts, attention and actions, and the orbital and ventromedial PFC (VMPFC) regulate emotions [12]. Functional Magnetic Resonance Imaging (fMRI) studies have reported the medial PFC in combination with the anterior cingulate cortex regulate human stress response during Stroop and Montreal Imaging Stress Tasks[13,14]. Another study reported that, stress reduced working memory-related dorsolateral prefrontal cortex (DLPFC) activity [9]. Up to date, only few neuroimaging studies have investigated the associations between stress and functional brain connectivity [15-18].

Functional connectivity (FC) is a mechanism that links brain regions sharing common functional properties [19]. FC could serve as a promising marker in the diagnosis and prediction of neuropsychiatric disorders illness and injuries [20-24]. FC usually investigated using modern neuroimaging modalities such as fMRI, EEG and MEG [25-27]. EEG and MEG have excellent temporal resolution enables them to study the coherency between brain region within minimum acquisition time. However, the MEG possess several limitations such as, cost, portability, and source localizations [28]. Functional near-infrared spectroscopy (fNIRS) is a novel and promising for cost effective and noninvasive brain imaging in research and clinical practice [29]. fNIRS measures local changes in hemoglobin concentrations within cortical layers through light sensors placed on the surface of the scalp [29]. Thus, it detects hemodynamic modulations as an indirect measure of neural activity similar to the blood oxygen level dependent (BOLD) functional magnetic resonance imaging (fMRI) signal. fNIRS has several advantages over other imaging techniques such as PET, SPECT and fMRI because of its portability, easy to administer, tolerate small movements, ability to perform long data acquisition and provide high temporal resolution [30]. Additionally, it has a good spatial resolution compare to EEG modality which give it the potential for widespread implementation [31].

To date, fNIRS has been increasingly used not only to localize focal brain activation during cognitive engagement [32], but also to map the functional connectivity of spontaneous brain activity during resting and task states in normal people and patient with psychiatric disorders [33-44]. fNIRS-based connectivity is a novel analysis tool for fNIRS data from the perspective of functional interactions which could be complementary to fNIRS activation analysis [45]. It is widely assumed that functional connections reflect neuroanatomical substrates [46]. Strong temporal correlations of spontaneous fluctuations of distinct regions in the brain are found in frequency range (<0.1 Hz) during resting state [47]. fNIRS has a higher time resolution (sampling rate: >10 Hz) than fMRI (sampling rate: ~1 Hz), which prevents aliasing of higher frequency activity such as respiratory (~0.15-0.3 Hz) and cardiovascular activity (~0.6-2 Hz) into low frequency signal fluctuations [48]. This results in more reliable estimation of functional connectivity at rest or active states. Studies using fNIRS demonstrated that spontaneous oscillations of cerebral hemodynamics include two distinguishable frequency components at low frequency (~0.1 Hz) and at very low frequency (~0.04 Hz) [33,49,50]. Although the mechanism underlying these signal fluctuations remains unknown, simultaneous recordings of cerebral hemoglobin oxygenation, heart rate, and mean arterial blood pressure showed that the systemic signal contribution to the hemodynamic changes in the frequency range (0.04–0.15 Hz) was 35% for $O_2Hb$ and 7% for HHb [51], suggesting that low-frequency fluctuations largely reflect hemodynamic responses to regional neural activities.

Based on fNIRS, functional connectivity at resting state has been demonstrated to exhibit distinct frequency-specific features [33,43,50,52,53]. For instance, Wu et al. demonstrated that the correlations among cortical networks are concentrated between ultralow frequencies of 0.01 Hz to 0.06 Hz [52]. Sasai et al. reported that functional connectivity between the homologous cortical regions of the contralateral hemisphere showed high coherence in the frequency range of 0.009 Hz to 0.1 Hz [33]. Medvedev et al. demonstrated hemispheric asymmetry within the functional architecture of the brain at low frequency range of 0.01 Hz to 0.1 Hz over the inferior frontal gyrus and the middle frontal gyrus [50]. Specifically, the study found a leading role of the right hemisphere in regard to left hemisphere as measured by granger causality. Xu et al. on the other hand, found reduced in coherence at frequencies ranges of 0.06Hz to 2 Hz and 0.052 Hz to 0.145 Hz in the PFC, and 0.021 Hz to 0.052 Hz in the motor cortex of healthy people at the end of driving task [43]. Another study demonstrated significant reduced connectivity in various frequencies intervals of 0.0095Hz to 0.021 Hz in homologous, 0.021 Hz to 0.052 Hz in front-posterior and 0.052Hz to 0.145 Hz in the motor-contralateral in subjects with cerebral infarction compared to that of healthy group [53]. This study aims to investigate the frequency-specific characteristics of fNIRS (narrow frequency band between 0.009 Hz to 0.02 Hz, and wide frequency band between 0.009 Hz to 0.1 Hz) and EEG (between 8 Hz to 13 Hz) over inter and intra-hemispheric PFC during active control and stress tasks. The study hypothesized that, stress affects the coherency of the entire frequency below 0.1 Hz and compared the results with well known EEG modality.

## 2. Materials and methods

### 2.1 Participants

Twenty-five male, young adults (aged 22 ± 4) participated in this study after signing consent form approved by the Universiti Teknologi PETRONAS review board. All subjects were right handed as determined by Edinburgh handedness questionnaire [54] and had normal or corrected-to-normal vision. They were in good health and had no history of neurological disorders, or psychotropic drug use and were not taking any type of medication at the time of experiment. Subjects were comfortably seated at distance of 80 cm from a 19-inch monitor in a quiet room and instructed to avoid head and body movements. They were also asked to fix gaze on a white cross on the black screen during the entire experiment. The experiment was conducted in accordance with the Declaration of Helsinki and all procedure were carried out with adequate understanding of the subjects.

### 2.2 Stress task

The task performed in this study was based on arithmetic task described in detail in [55]. The task designed and presented in MATLAB with Graphical User Interface (GUI). It involved 3-single digit integer (from 0 to 9) and used the addition (+) or subtraction (–) operands (for example 2-3+9). The answers were presented in the sequence of 0 to 9 and participant has select the right answer by doing single-click using the mouse. The task was performed in three phases; training, control and stress phase, each of the phase has its own conditions. During the training phase, participants practiced the arithmetic task for a total duration time of 5 minutes. At this phase, the percentage of answering the task correctly and the time taken by each individual to answer each question was recorded. Subjects with poor performance were not counted for this study. Then, the average recorded time was reduced by 10% and used as time pressure to stress the participants. During the control phase, EEG+fNIRS probe holder was attached to the participant's head and signals were simultaneously recorded for a total duration of 5 minutes. At this phase, each of the participants was instructed to solve the arithmetic problems as fast as he can but without any time limit per question. At the end of the control phase, participants filled up a self-report questionnaire about workload using NASA-TLX [56] described in the next sub-section. During the stress phase, simultaneous measurement of EEG+fNIRS was recorded for a total duration of 5 minutes. At this phase, each of the participant solves the arithmetic problems under time pressure (the average time taken in the training phase which reduced by 10%). Besides the time pressure, feedback of answering the questions ("correct", "incorrect" or "timeout") were displayed on the computer monitor to further induce stress in the participants. Additionally, performance indicators (one for the participant's performance and one for the averaged peer performance fixed at 90% accuracy) were shown on the monitor to induce more negative emotional stress. At the end of the stress phase, participants filled up another NASA-TLX self-report questionnaire about the task loading.

Figure 1 summarizes the overall experimental procedures. A block design that incorporated both; the control and the stress tasks was used in the presented study. A total of 10 blocks were used (5 blocks in the control phase and 5 blocks in the stress phase). In both phases; control and stress, the arithmetic tasks in each block were display for 30 seconds followed by 20 seconds rest. During the 20 s rest, participants instructed to focus on a fixation cross with a black background to sustain their attention to the monitor display. The order of the task conditions was presented randomly. The reaction time and the accuracies of answering the task under the control and the stress phases were recorded and used for analysis of behavioural data.

All participants report they were relaxed and peaceful during the control-workload and were stressed while performing the task under stress condition.

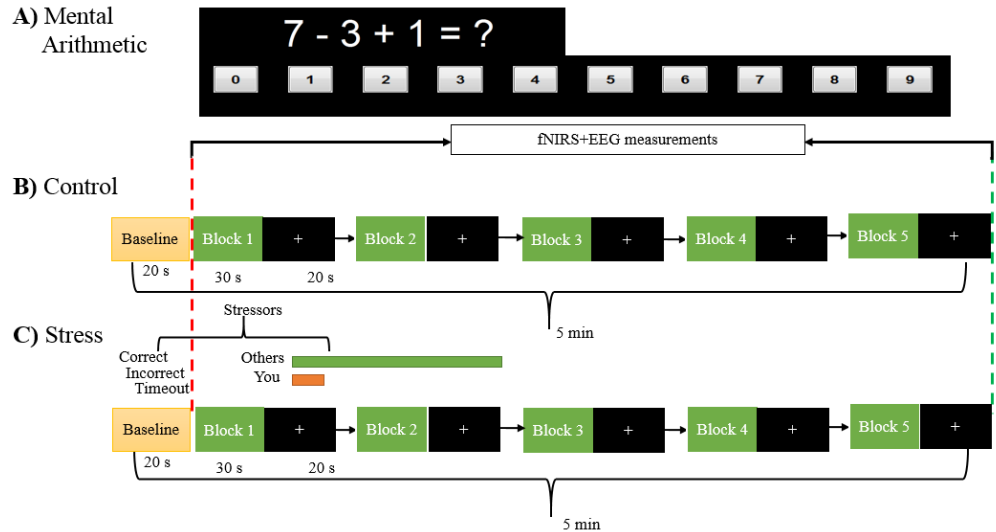

Fig. 1. Schematic of the experimental procedure. (A) The presentation of the arithmetic task. (B) Blocks design of the simultaneous measurements under control condition. (C) Block design of simultaneous measurement under stress condition. In both; the control and stress conditions, there were ten blocks. Mental arithmetic in each block was allocated for 30 s followed by 20 s rest.

*2.3 Simultaneous measurement of EEG+fNIRS*

EEG and fNIRS data were simultaneously acquired from the PFC area. The EEG electrodes were integrated with fNIRS opteds in one single probe holder. EEG data were recorded continuously using BrainMaster 24E system. Seven-active EEG electrodes that covers the PFC (FP1, FP2, F3, F4, Fz, F7, and F8) and two ground electrodes placed on earlobes A1+A2 were used in the measurements. The EEG data were sampled at a frequency of 256 Hz and their impedances were maintained below 5kΩ. On the other hand, the hemodynamic responses were recorded using optical topography system (OT-R40, Hitachi Medical Corporation, Japan). The fNIRS optodes were placed on the surface above the PFC area guided by the EEG electrode positions.

The light sources in fNIRS consisted of continuous laser diodes with two wavelengths, 695 nm, and 830 nm. The transmitted light was sampled every 100 ms. The fNIRS system was equipped with 16 optical fibres; 8-sources and 8-detectors placed over the PFC area between locations FP1/2-F3/4-F7/8 thus covering the Frontopolar area (FPA, CH 9-11, 15, 16, 20-22), Dorsolateral (DLPFC; 1-7) and Ventrolateral (VLPFC; 8, 13, 14, 19, 12, 17, 18, and 23). These sources and detectors were connected to the subject's head by flexible optical fiber bundles using the anatomical landmarks provided by the international 10-20 system. The inter-optode distance was 30 mm, which allowed for measuring neural activities approximately 15-25 mm beneath the scalp. The optodes distribution over the PFC is as demonstrated in Fig. 2(A). The measurement area between a pair of source-detector probes was defined as a channel (CH). As a result, 23 optical channels were recorded. The arrangement and appearance of the probe array/electrode with overall experiment set up are shown in Fig 2. We controlled the simultaneous measurement of EEG+fNIRS using MATLAB. Two triggers were sent, one through the serial port and the other one through the parallel ports to mark the start and the end of the task in each block of the mental arithmetic task in the fNIRS and EEG systems respectively.

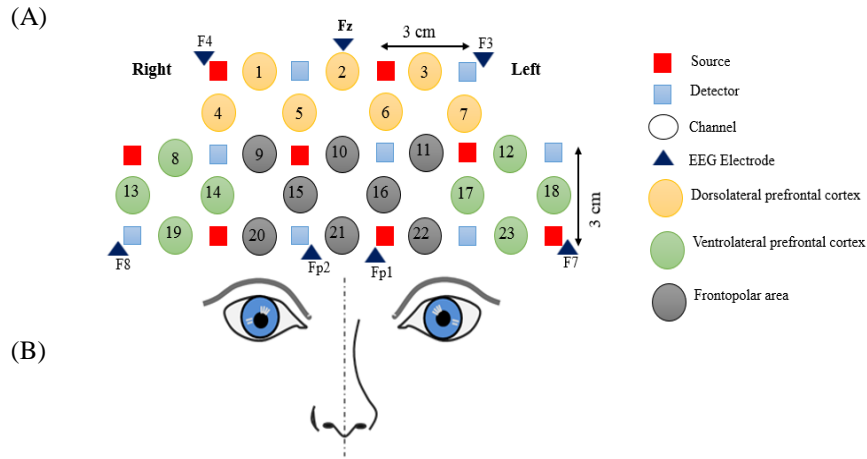

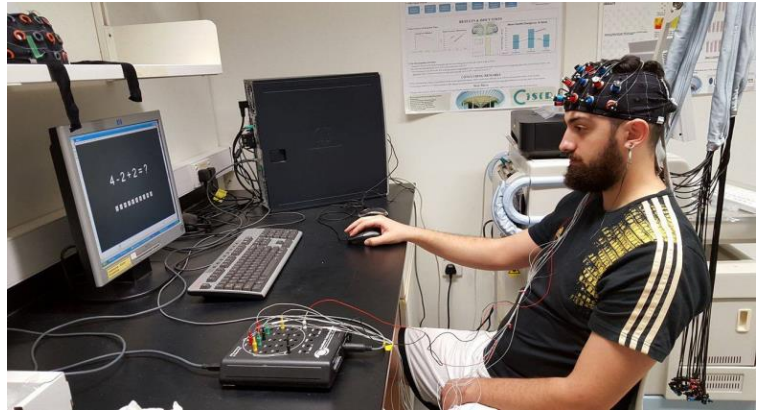

Fig. 2. Probe setting and schematic arrangement points. (A) Location of the fNIRS probe array/optodes (red=sources; blue= detectors) and EEG electrodes according to the international 10-20 system. Channels (numbered from 1 to 23) were measured at three lateral PFC subregion. Each subregion/anatomical area represented by different color. There were total of twenty-three fNIRS channels and 7-EEG electrode. (B) Overall experiment set-up and task presentation.

## *2.4 Data processing*

Both data were preprocessed and analyzed offline, separately. EEG data were preprocessed using EEGLAB toolbox version 9.0.4 with custom scrip [57]. To consider the highly correlated frequency band responses to mental stress, the data were band-filtered between 8 Hz and 13 Hz using Butterworth filter [55]. Additionally, the low-frequency drift and high frequency noise were eliminated at this frequency band. The eye-blinks and movement artefacts were removed manually and using Independent component analysis (ICA) technique. Each of the EEG channel was decomposed into N number of components, and the component described prefrontal eye blink artifacts was manually rejected. Baseline correction was performed by subtracting the average value of the first 200ms rest state from the signal the EEG signals at all subsequent data points. Due to the effects of referencing on the functional connectivity on EEG signals, all electrodes were re-referenced to the average linked-earlobes (A1+A2)[58].

For fNIRS data, the processing and analysis was performed using custom script as well as the developed plug-in analysis software Platform for Optical Topography Analysis Tool [59]. The raw fNIRS data were transformed to the concentration changes of oxygenated and de-oxygenated hemoglobin using modified Beer-Lambert approach. The signals of the $\Delta O_2Hb$ and $\Delta HHb$ were then band-pass filtered with 0.009 Hz and 0.1 Hz to remove the long-term drift of baseline, Mayer waves, high frequency cardiac and respiratory activities [60,61]. This is due to that, studies have shown that fNIRS signals like other blood-related brain measures are low frequency oscillations detectable mainly between 0.01 Hz and 0.1 Hz [62]. A period from the onset of the task condition to the end period of the task condition (30 s) was defined as one single analysis block. Each data block was baseline-corrected by subtracting the mean value of the fNIRS data measured at the pre-task period from each data point in the task period [63]. After baseline correction, the five blocks were concatenated to form a continuous signal of 2.5 minutes (Five blocks ×30 s). The functional connectivity was then measured from the concatenated signals at significant 2.5 minutes duration time [64].

*2.5 functional connectivity based on coherence*

The effects of mental stress on inter and intra-hemispheric functional connectivity is measured by calculating the magnitude squared coherence. We used Welch's method to compute a modified periodogram technique (using a 1024-point Fourier transform, Hamming window, and 512-point overlap) [65] to estimate the cross-spectral and power spectral density of the signals [33]. The coherence measures the linear time-invariant relationship between two time series signals, $x$ and $y$ at frequency $f$ [66,67]. The coherence is calculated as:

$$Coh_{xy}(\lambda) = \frac{|f_{xy}(f)|^2}{f_{xx}(f)f_{yy}(f)} \quad , \quad (1)$$

where $f_{xy}(f)$ is the cross-spectrum of $x$ and $y$, and $f_{xx}(f)$ is the power spectrum of x. The coherence is a positive function bounded by 0 and 1, where 0 indicates that $x$ can perfectly predict $y$ in a linear fashion. The functional connectivity was then mapped based on squared coherence threshold of >0.6.

For EEG, the FC is investigated at alpha frequency band signals (range between: f= 8-13 Hz). The FC within inter-hemispheric is based on electrode pairs of: F4↔F3, F8↔F7, and FP2↔FP1. Similarly, the FC within intra-hemispheric PFC; right hemisphere is based on pairs of; FP2↔F8, FP2↔F4, F4↔F8 and the left hemisphere; FP1↔F3, FP1↔F7, F7↔F3, respectively.

For fNIRS, the FC is investigated at two frequency intervals; (range between; $f_1$=0.009-0.02 Hz; and $f_2$=0.009- 0.1 Hz) which encompasses the frequencies of hemodynamic responses at $O_2Hb$ signals [33]. The selection of these frequency intervals is due to that, the typical frequency band used in the fMRI field to assess functional connectivity is in the range of 0.01 Hz to 0.1 Hz because other bands are contaminated by noise and physiological artifacts such as respiratory and cardiac-related fluctuations in oxygen supply [68].

The FC within inter hemispheric is based on channel pairs of; CH1↔CH3, CH4↔CH7, CH5↔CH6, CH8↔CH12, CH9↔CH11, CH13↔CH18, CH14↔CH17, CH15↔CH16, CH19↔CH23 and CH20↔CH22. Similarly, the FC within right hemispheric is based on channel pairs between all channels; CH1, CH4, CH5, CH8, CH9, CH13-, CH14, CH15, CH19, CH20 and the FC within the left hemispheric is based on channel pairs between all of; CH3, CH6, CH7, CH11, CH12, CH16, CH17, CH18, CH22, CH23, respectively. This indicates that, for perfect connectivity, each channel should be bounded to nine channels from the same intra-hemispheric. The FC is mapped using different colors to indicate the various connectivity strengths in the coherence maps of EEG and fNIRS responses. To better understand the effects of stress on PFC subregion-FC, all electrode/Channel pairs in inter and intra-hemispheric PFC are classified into three connection groups: (1) Connectivity between the DLPFC regions, (2) Connectivity between the VLPFC regions; and (3) Connectivity between the frontopolar areas.

*2.6 Statistical analysis*

To evaluate the effects of stress on the prefrontal functional connectivity within the inter and intra hemispheric, we performed statistical analysis using t-test. The t-test analysis was performed to measure the differences between the control and stress group in the subjective measurements as well as the EEG measured coherence values. For fNIRS, the coherence values of the narrow frequency band and the wide frequency band were averaged in the control group as well as the stress group and then we applied the t-test on the average differences between the control group and stress group. For EEG and fNIRS, we analyzed the differences between them in electrode /channel basis criteria. In each electrode/channel, the differences in functional connectivity were considered statistically significant if the t-value is greater than or equal to 3 ($t \geq 3$); corresponding to p-value of less than 0.01, $p < 0.01$.

## 3. Result and analysis

*3.1 Self-report evaluation*

Self-report evaluation is very important physiological measure of mental stress. The self-report questionnaire reveals that participants were not stressed before they perform the task under time pressure with negative feedback and were stressed after they perform the task. According to the result obtained from NASA-TLX subscale, there was a significant increase in the scores of mental demands (MD, $p<0.001$), physical demand (PD, $p<0.002$), temporal demand (TD, $p<0.002$), performance (PF, $p<0.0001$), effort (EF, $p<0.0001$) and frustration (FR, $p<0.0005$) respectively. This indicates that the task induced moderate level of stress and much effort is required to maintain the performance at its highest level.

*3.2 Inter-hemispheric PFC connectivity*

The EEG functional connectivity results of the inter-hemispheric PFC under neutral-control condition and under stress condition are shown in Fig.3 (a) and Fig.3 (b), respectively. The connectivity between brain regions is represented by their coherence maps with varies strengths. Under neutral-control condition, the results show that brain regions are highly bounded/connected especially at the bilateral nodes. Additionally, connectivity at the position of Fp2 and F7 demonstrated the highest strength compare to other nodes. Under stress condition weak connectivity is observed specifically within the DLPFC and at the nodes that links the DLPFC to the FPA and DLPFC to the right VLPFC respectively.

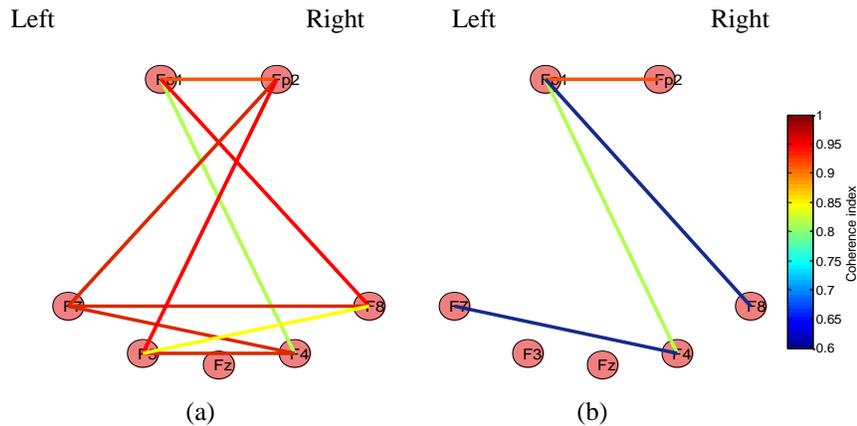

Fig. 3. EEG coherence map connectivity of the inter-hemispheric PFC areas at seven electrode positions. (a) EEG connectivity under neutral-control condition and (b) EEG connectivity under stress condition. The map represents the coherence index value with varies strength. Blue color represents less connectivity and dark-red color represent higher connectivity map across EEG nodes.

The fNIRS functional connectivity results of inter-hemispheric PFC on $O_2Hb$ under neutral-control and under stress condition at narrow frequency band of 0.009- 0.02 Hz are shown in Fig.4 (a) and Fig.4 (b) respectively. Similarly, the FC at wider frequency band of 0.009- 0.1 Hz is shown in Fig.4 (c, d) for the control and stress group respectively. The topographical maps in Fig.4 represent the connectivity between the right and left nodes of the PFC with strength varies according to their coherency. From Fig.4 (a, c) it's clearly seen that, the control task enhances the overall connectivity between the right and left hemispheric nodes with highest strength at the FPA and VLPFC areas. It's also observe that the functional connectivity obtained at narrow frequency band of 0.009-0. 02 Hz can be fully estimate at wider frequency of 0.009- 0.1 Hz. This indicates that, the $O_2Hb$ is less affected by noise at frequency below 0.1 Hz. From Fig.4 (b, d) it's clearly seen that, stressful task reduces the overall connectivity between the right and left hemisphere with much reduction over the DLPFC areas. The reduced in the FC under stress is observed within narrow and wide-band frequency intervals but not equally as seen in the control group. Under stress condition, it's observed that, the FC reduced albeit with wider frequency band but not significant. From the overall results, it is clearly seen that, the functional connectivity obtained at narrow frequency band (between 0.009-0.0002 Hz) can be obtained at wider frequency band (0.009-0.1 Hz) with similar strengths as demonstrated by their coherence maps under the neutral-control and stress conditions. This confirms that, the fNIRS technique is highly resistant to movement artefacts and noise at frequency below 0.1 Hz.

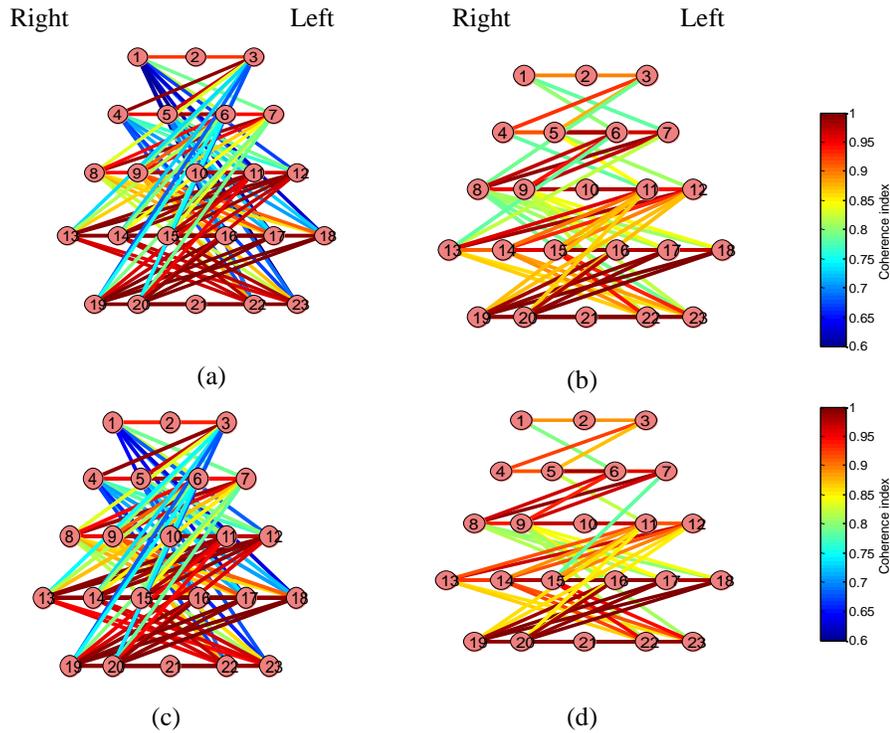

Fig. 4. fNIRS coherence map of inter-hemispheric functional connectivity at narrow frequency band (0.009-0.02 Hz); (a) $O_2Hb$ connectivity under neutral-control condition and (b) $O_2Hb$ connectivity under stress condition and wider frequency band(0.009-0.1 Hz); (c) $O_2Hb$ connectivity map under neutral-control condition and (d) $O_2Hb$ functional connectivity under stress condition. The strength of the connectivity varies as represent by colors; blue color reveals less connectivity and dark-red color reveals higher connectivity map across the fNIRS channels/nodes.

Similarly, the functional connectivity results of inter-hemispheric PFC on HHb under neutral-control and under stress condition at narrow frequency band of 0.009- 0.02 Hz are shown in Fig.5 (a) and Fig.5 (b) respectively. At wider frequency band of 0.009- 0.1 Hz the FC is shown in Fig.5(c, d) for the control and stress group respectively. Obviously, there is a decrease in the functional connectivity from the control condition to the stress condition in both frequency intervals. It is also observed that, the FC obtained at narrow frequency band can be obtained within a wider frequency band. Interestingly, there is a decrease in the FC between brain region under the control condition as well as under the stress condition compare to the functional connectivity obtained with $O_2Hb$. This confirms the stability of $O_2Hb$ concentration and suggest it as an excellent metric to obtain functional connectivity between brain regions.

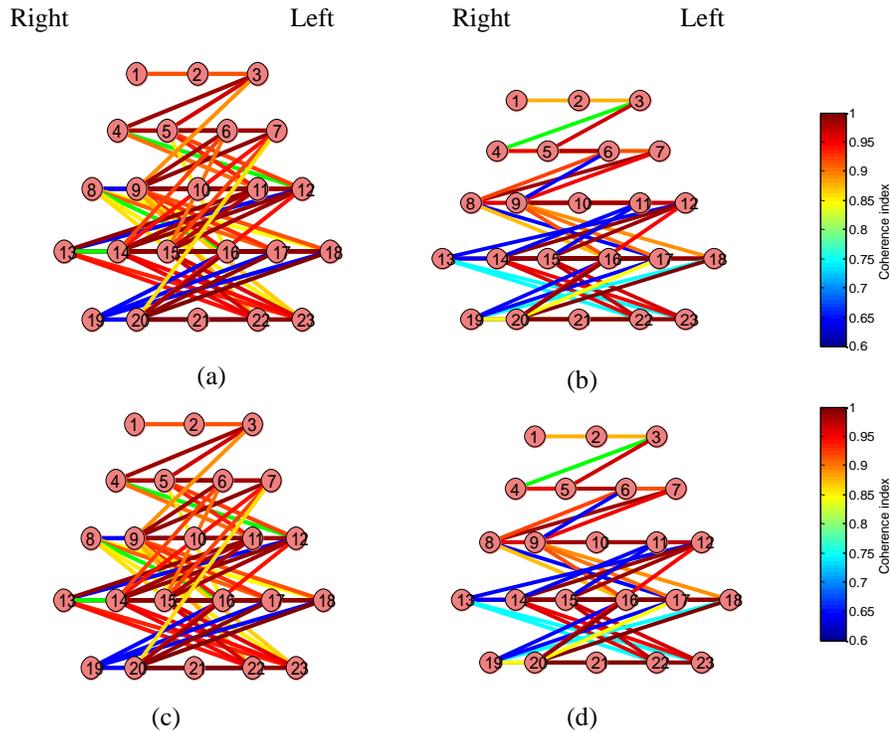

Fig. 5. fNIRS coherence map of inter-hemispheric functional connectivity at narrow frequency band (0.009-0.02 Hz); (a) HHb connectivity under neutral-control condition and (b) HHb connectivity under stress condition and wider frequency band(0.009-0.1 Hz); (c) HHb connectivity map under neutral-control condition and (d) HHb functional connectivity under stress condition. The strength of the connectivity varies as represent by colors; blue color reveals less connectivity and dark-red color reveals higher connectivity map across the fNIRS channels/nodes.

### 3.3 Intra-hemispheric PFC connectivity

The EEG functional connectivity on the intra-hemispheric PFC areas is shown in Fig.6 (a) under neutral-control and Fig.6 (b) under stress condition respectively. The topographical maps show that, the connectivity is reduced with varies strengths from control to stress condition over the FPA. The reduced in connectivity under stress indicates that, the stress not only affects the global/entire network but also the within subregion-specific network.

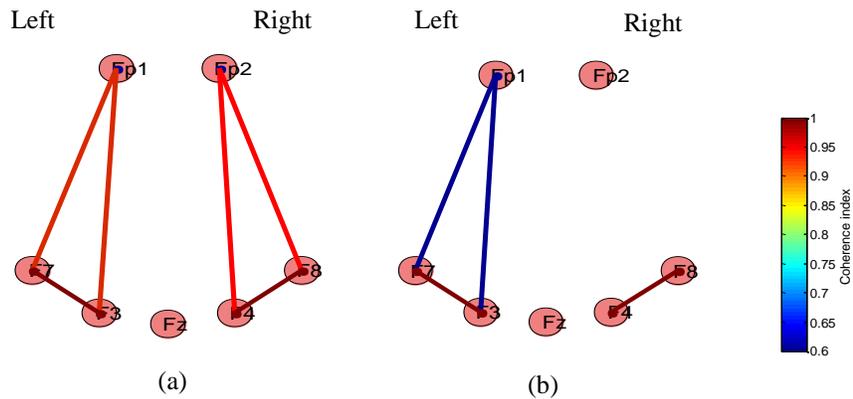

Fig. 6. EEG coherence map connectivity of the intra-hemispheric PFC areas at seven electrode positions. (a) EEG connectivity under neutral-control condition and (b) EEG connectivity under stress condition. The map represents the coherence index value with varies strength. Blue color represents less connectivity and dark-red color represent higher connectivity map across EEG nodes.

The results of intra-hemispheric functional connectivity based $O_2Hb$ within the narrow frequency band of 0.009-0.02 Hz and wide frequency interval of 0.009-0.1 Hz under the neutral-control and stress condition are shown by their topographical coherence maps in Fig.7(a, b) and Fig.7(c, d) respectively. The topographical maps show that the connectivity decrease from control to stress in the intra-hemispheric within the two frequency intervals with almost equal strength. It's also observed that, the strength of the connectivity between the DLPFC areas is less than the strength in the other PFC subregions. This supports the focality of the hemodynamic under stress condition to a specific subregion, in this case the dorsolateral PFC area.

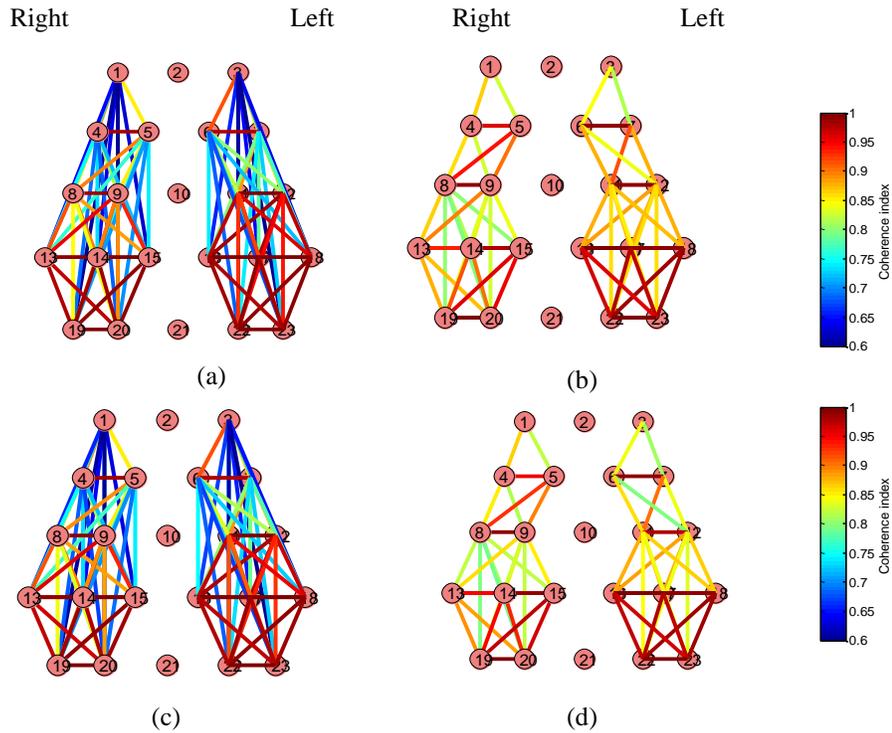

Fig. 7. fNIRS coherence map of inttra-hemispheric functional connectivity at narrow frequency band (0.009-0.02 Hz); (a) O$_2$Hb connectivity under neutral-control condition and (b) O2Hb connectivity under stress condition and wider frequency band(0.009-0.1 Hz); (c) O2Hb connectivity map under neutral-control condition and (d) O$_2$Hb functional connectivity under stress condition. The strength of the connectivity varies as represent by colors; blue color reveals less connectivity and dark-red color reveals higher connectivity map across the fNIRS channels/nodes.

Figure 8 shows the results of functional connectivity of the intra-hemispheric obtained from the HHb at narrow and wide frequency intervals under neutral-control, Fig.8 (a, c) and stress condition, Fig.8 (b, d), respectively. The topographical maps show a decrease in the functional connectivity from control condition to stress condition in the DLPFC areas. Additionally, there is albeit reduce in the connectivity under stress when increasing the frequency intervals from (0.009- 0.02 Hz to 0.00-0.1 Hz) but not significant.

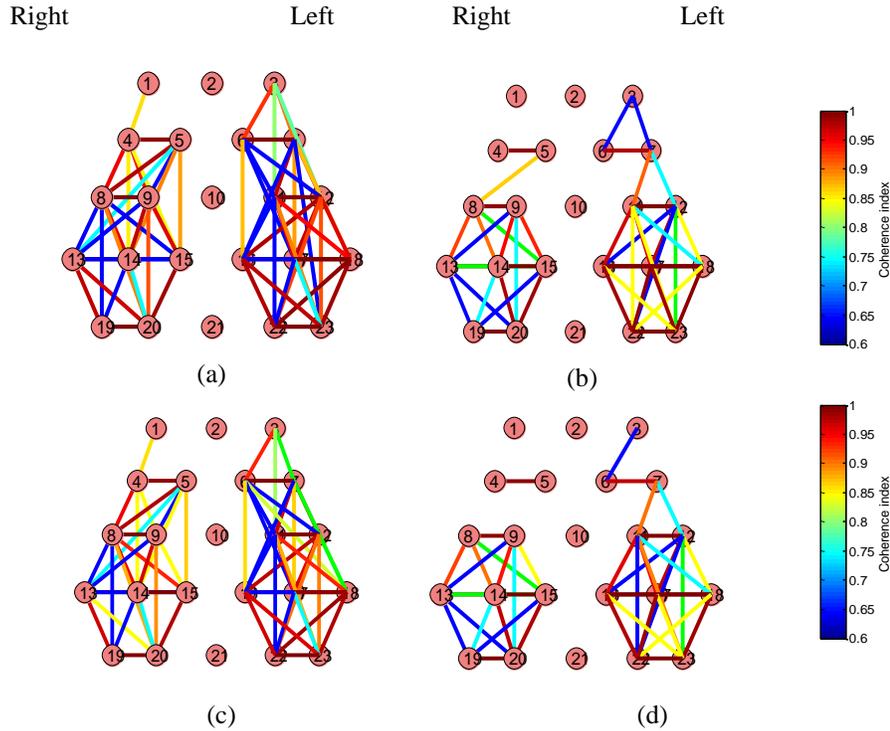

Fig. 8. fNIRS coherence map of inter-hemispheric functional connectivity at narrow frequency band (0.009-0.02 Hz); (a) HHb connectivity under neutral-control condition and (b) HHb connectivity under stress condition and wider frequency band (0.009-0.1 Hz); (c) HHb connectivity map under neutral-control condition and (d) HHb functional connectivity under stress condition. The strength of the connectivity varies as represent by colors; blue color reveals less connectivity and dark-red color reveals higher connectivity map across the fNIRS channels/nodes.

*3.4 Statistical analysis results*

The results of the statistical analysis demonstrated significantly reduce in overall functional connectivity within specific locations on the PFC areas. Taking each individual EEG-electrode/ fNIRS-channel, the average t-values and p-values for the average of two frequency intervals in all the participants are summarized in Table 1 based on the $O_2Hb$ for inter and intra-hemispheric and Table 2 based on HHb for inter and intra-hemispheric between the control and stress conditions. Considering thresholding value of $t \geq 3$ as discussed in section 2.6, the effects of stress are highly localized to specific subregions on the PFC area as given in Table1 and Table 2. For EEG, only two electrodes FP2, F3 on the inter-hemispheric PFC one electrode from the intra-hemispheric PFC area respond above the given threshold value, respectively. For the $O_2Hb$ six channels from the dorsolateral; CH1-CH7 on the inter-hemispheric PFC area and nine channels CH1-CH7, CH15 and CH19 on the intra-hemispheric PFC area respond above the given thresholding value respectively. For the HHb three channels; CH11, CH12, CH14 on the inter-hemispheric PFC area and seven channels CH1-CH6, and CH14 on the intra-hemispheric PFC area respond above the given thresholding value respectively. Taking all together, inter and intra-hemispheric PFC, the functional connectivity results demonstrated that, the dorsolateral and right-ventrolateral PFC area are the most sensitive subregions to mental stress.

**Table 1. Statistical analysis of functional connectivity of the inter and intra-hemispheric PFC based on O2Hb Means ± SE between control and stress conditions**

| | Inter-hemisphere (O2Hb) | | | Intra-hemisphere (O2Hb) | |
|---|---|---|---|---|---|
| CH | t-value | p-value | CH | t-value | p-value |
| 1 | 4.95±0.3 | 0.0001 | 1 | 4.55±0.3 | 0.0001 |
| 3 | 3.83±0.2 | 0.0001 | 3 | 4.3±0.2 | 0.0001 |
| 4 | 4.80±0.2 | 0.0001 | 4 | 4.40±0.2 | 0.0001 |
| 5 | 3.51±0.2 | 0.0010 | 5 | 3.71±0.2 | 0.0001 |
| 6 | 3.35±0.2 | 0.0040 | 6 | 3.85±0.2 | 0.0005 |
| 7 | 3.30±0.3 | 0.0010 | 7 | 3.50±0.3 | 0.0005 |
| 8 | 1.24±0.4 | 0.2120 | 8 | 0.84±0.2 | 0.2122 |
| 9 | 1.70±0.4 | 0.1310 | 9 | 0.87±0.4 | 0.2123 |
| 11 | 1.90±0.4 | 0.1310 | 11 | 1.10±0.3 | 0.2100 |
| 12 | 1.93±0.4 | 0.1300 | 12 | 0.93±0.4 | 0.4342 |
| 13 | 2.81±0.1 | 0.0110 | 13 | 2.21±0.1 | 0.0411 |
| 14 | 2.47±0.1 | 0.0251 | 14 | 0.87±0.2 | 0.2211 |
| 15 | 2.90±0.2 | 0.0062 | 15 | 3.10±0.2 | 0.0050 |
| 16 | 2.30±0.3 | 0.0510 | 16 | 2.10±0.3 | 0.0488 |
| 17 | 2.83±0.3 | 0.0130 | 17 | 0.93±0.3 | 0.1921 |
| 18 | 2.87±0.2 | 0.0100 | 18 | 2.57±0.2 | 0.0231 |
| 19 | 2.92±0.1 | 0.0101 | 19 | 3.67±0.1 | 0.0005 |
| 20 | 2.73±0.3 | 0.0102 | 20 | 2.76±0.2 | 0.0111 |
| 22 | 1.95±0.2 | 0.1212 | 22 | 2.55±0.2 | 0.0169 |
| 23 | 2.86±0.2 | 0.0111 | 23 | 2.86±0.2 | 0.0111 |
| Fp1 | 0.91±0.2 | 0.2222 | Fp1 | 0.91±0.3 | 0.2000 |
| Fp2 | 3.52±0.1 | 0.0010 | Fp2 | 4.57±0.2 | 0.0001 |
| F3 | 4.73±0.2 | 0.0001 | F3 | 1.68±0.1 | 0.1111 |
| F4 | 1.95±0.2 | 0.1211 | F4 | 2.66±0.2 | 0.0121 |
| F7 | 2.89±0.2 | 0.0100 | F7 | 0.85±0.2 | 0.2121 |
| F8 | 2.91±0.2 | 0.0100 | F8 | 2.76±0.1 | 0.0111 |

**Table 2. Statistical analysis of functional connectivity of the inter and intra-hemispheric PFC based on HHb Means ± SE between control and stress conditions**

| | Inter-hemisphere (HHb) | | | Intra-hemisphere (HHb) | |
|---|---|---|---|---|---|
| CH | t-value | p-value | CH | t-value | p-value |
| 1 | 0.85±0.2 | 0.3100 | 1 | 3.95±0.10 | 0.0004 |
| 3 | 1.98±0.3 | 0.0910 | 3 | 3.10±0.11 | 0.0052 |
| 4 | 2.80±0.2 | 0.0130 | 4 | 4.80±0.10 | 0.0002 |
| 5 | 2.90±0.2 | 0.0114 | 5 | 4.50±0.11 | 0.0001 |
| 6 | 1.90±0.2 | 0.3310 | 6 | 3.90±0.13 | 0.0005 |
| 7 | 2.90±0.2 | 0.0121 | 7 | 2.80±0.24 | 0.0112 |
| 8 | 0.93±0.2 | 0.3002 | 8 | 3.30±0.13 | 0.0023 |
| 9 | 2.90±0.2 | 0.0180 | 9 | 2.80±0.21 | 0.0142 |
| 11 | 3.14±0.3 | 0.0051 | 11 | 2.14±0.21 | 0.0491 |
| 12 | 3.16±0.3 | 0.0051 | 12 | 2.46±0.20 | 0.0226 |
| 13 | 0.94±0.1 | 0.3103 | 13 | 0.94±0.21 | 0.3004 |
| 14 | 3.12±0.1 | 0.0050 | 14 | 3.10±0.21 | 0.0052 |
| 15 | 2.81±0.2 | 0.0130 | 15 | 0.91±0.21 | 0.3007 |
| 16 | 0.87±0.3 | 0.3010 | 16 | 1.87±0.22 | 0.1114 |
| 17 | 0.86±0.3 | 0.3110 | 17 | 1.86±0.21 | 0.1211 |
| 18 | 0.87±0.2 | 0.3320 | 18 | 1.87±0.22 | 0.1113 |
| 19 | 0.83±0.1 | 0.4101 | 19 | 1.10±0.10 | 0.4124 |
| 20 | 0.85±0.2 | 0.4011 | 20 | 1.82±0.21 | 0.1212 |
| 22 | 0.85±0.2 | 0.4011 | 22 | 1.85±0.22 | 0.1210 |
| 23 | 2.60±0.2 | 0.0169 | 23 | 1.36±0.20 | 0.2322 |

## 4. Discussion

This study investigated the effects of mental stress on the functional connectivity within the inter and intra-hemispheric based on synchronized electrical and hemodynamic responses of the brain. The electrical responses were investigated using alpha rhythmic fluctuations within the frequency range of 8-13 Hz. Similarly, the hemodynamic responses of oxygenated and deoxygenated hemoglobin were investigated with narrow frequency interval of 0.009 Hz to 0.02 Hz and with wider frequency interval of 0.009 Hz to 0.1 Hz. The stress demonstrated significantly reduce in functional connectivity on specific sub-region based on the synchronized electrical and hemodynamic responses.

The task at control condition improves the functional connectivity across inter and intra-hemispheric PFC areas based on the patterns of the EEG and fNIRS signals. The interactions between PFC sub-regions were highly bounded within the dorsolateral followed by that at the ventrolateral and frontopolar areas while performing the task under neutral-control condition. Although, the coherence index cannot show the direction of interactions, the results showed that all brain regions were connected with higher strength at the frontopolar and ventrolateral areas in both; inter and intra-hemispheric PFC. Taking each hemisphere separately, the left hemisphere was highly bounded in the EEG rhythm as well as in the $O_2Hb$ and HHb compare right hemisphere under neutral-control condition. This phenomena could be due to the effects of mental arithmetic execution which has a left dominant as reported in previous studies [55,69]. Note that, the results of the functional connectivity in this study on $O_2Hb$ is highly connected compare to that of HHb. This is due to that, $O_2Hb$ has better signal-to-noise ratio (SNR) than HHb and had proven as a better indicator of changes in regional cerebral blood flow and highly correlated with fMRI BOLD signals [70,71].

On the other hand, the task with time pressure and negative feedback reduced the overall functional connectivity from neutral-control to stress condition within inter and intra-hemispheric PFC subregions in the EEG rhythmic as well as in the $O_2Hb$ and HHb respectively. Specifically, the functional connectivity at the right-dorsolateral area was highly affected by stress in both; inter and intra-hemispheric and the two rhythmic intervals of $O_2Hb$ and HHb. The results of $O_2Hb$ and HHb showed that the reduced interactions in the brain network under stress is not subserved by rhythmic neural synchronization at frequency below 0.1 Hz on the hemodynamic responses. In summary, the results of this study demonstrated that EEG and fNIRS-based functional connectivity reveals different patterns for different mental states (neutral-control and stress) in all the frequency intervals. This recommends the functional connectivity within brain sites as good indices to discriminate the stress state from that of control state.

## 5. Conclusion

This study investigated the effects of mental stress on PFC functional connectivity based on synchronized electrical alpha-rhythmic and hemodynamic responses of oxygenated and deoxygenated hemoglobin. The results showed that the functional connectivity significantly reduced over the dorsolateral PFC during stress condition as compared to neutral-control. The results of fNIRS functional connectivity based on oxygenated and deoxygenated hemoglobin showed that, stress affects the inter as well as intra-hemispheric PFC with entire frequency of below 0.1 Hz with no difference in connectivity measures on wide and narrow frequency bands. This study clearly shows the potential of the EEG and fNIRS modalities to measure the effects of stress on functional connectivity of the PFC and report the connectivity over the dorsolateral PFC mostly disrupted by stress.

## Acknowledgments

The author would like to thank all the subjects participated in the experiment for their patience during the EEG-fNIRS recording.